\documentclass[journal=nalefd,manuscript=letter,layout=twocolumn]{achemso}

\usepackage{amsmath,amssymb}
\usepackage{graphicx}
\usepackage{bm}
\usepackage{hyperref}

\title{Nonlocal Electrostatic Origin of Schottky-Barrier Variability in 2D Contacts}

\author{Hangbo Zhou}
\affiliation{Institute of High Performance Computing (IHPC), Agency for Science, Technology and Research (A*STAR), 1 Fusionopolis Way, \#16-16 Connexis, Singapore, 138632, Republic of Singapore}

\author{Yong-Wei Zhang}
\email{zhangyw@a-star.edu.sg}
\affiliation{Institute of High Performance Computing (IHPC), Agency for Science, Technology and Research (A*STAR), 1 Fusionopolis Way, \#16-16 Connexis, Singapore, 138632, Republic of Singapore}

\date{\today}


\begin{document}

\begin{abstract}
Electrical contacts often limit the performance of atomically thin semiconductor devices. The Schottky barrier height (SBH) is conventionally treated as a local interface property, yet reported values for the same metal/2D-semiconductor contact vary by hundreds of meV. Here we show that, in top contacts, the effective SBH exhibits a pronounced nonlocal electrostatic dependence on defects near the contact edge, beyond the conventional local interface framework. A nonlocal electrostatic model, supported by density-functional-theory-based transport calculations for Ti--MoS$_2$ and Au--MoS$_2$, captures the large, metal-dependent variations in SBH as a function of defect position relative to the contact edge. These results provide a unified explanation for the longstanding variability in experimentally extracted SBHs and establish nonlocal electrostatics, mediated by edge-proximal defects, as a key mechanism governing carrier injection in 2D contacts.
\end{abstract}

\maketitle

\par\vspace{1.5\baselineskip}

Control of the Schottky barrier height (SBH) is central to 2D electronics because contacts often limit carrier injection in atomically thin transistors~\cite{Li2023_Nature,Shen2021_Nature,Liu2021_Nature,Allain2015}. In transition metal dichalcogenides, with monolayer MoS$_2$ as a prototypical platform~\cite{Mak2010_PRL,Wang2012_NatNano}, even modest barrier variations can strongly affect contact resistance, on-state current, and device-to-device reproducibility~\cite{Allain2015,Wang2021_RPP}. This sensitivity is part of the broader effort to engineer low-resistance, tunable 2D contacts, including semimetal contacts, quantum-contact architectures, van der Waals interfaces, edge-engineered geometries, two-dimensional metallic electrodes, and high-work-function p-type contacts~\cite{Shen2021_Nature,Li2023_Nature,Liu2018_Nature,Wang2019_NatureVdWContacts,Zheng2019_NatElectron,Song2020_NatElectron,Wang2022_NaturePType}. A persistent difficulty, however, is that extracted SBHs for the same metal--semiconductor pair still differ by hundreds of meV. For MoS$_2$, reported values span about $0.05$--$0.95$ eV for Au contacts and about $0.05$--$0.46$ eV for Ti contacts~\cite{Liu2018_Nature,Qiu2012_APL,Das2013_NanoLett,Kaushik2014_APL,Kwak2014_NanoLett,Kim2017_ACSNano,Li2020_Nanotechnology,Xie2022_Nanotechnology}. The origin of this wide spread remains unresolved.

The standard theoretical description of SBH is rooted in local interface physics. Departures from the Schottky--Mott limit are commonly attributed to interface gap states, whether induced by the metal or by defects, which establish charge neutrality and produce Fermi-level pinning~\cite{Tung2000_PRL,Gong2014_NanoLett,Sorkin2022_SciRep,Sorkin2025_PCCP}. Within that picture, local perturbations are expected to be screened by the interfacial charge reservoir. In parallel, the pinch-off picture of inhomogeneous Schottky contacts shows that low-barrier regions do not inject independently. Their electrostatics are constrained by the surrounding higher-barrier background, so transport is governed by an effective saddle-point barrier~\cite{Tung2001_PRB,TungInhomSchottky}. Together, these ideas suggest that microscopic disorder should be substantially averaged out. The experimentally reported giant spread is therefore difficult to reconcile with a picture in which the relevant SBH is determined only by local interface properties.

Low-dimensional contacts already suggest that the experimentally relevant injection barrier can be shaped by geometry as well as by local chemistry~\cite{Wang2013_Science,Leonard2000_PRL}. This point is especially important for top contacts in 2D semiconductors, where carrier injection is governed by the contact edge rather than by a laterally averaged interface. Chalcogen vacancies and related point defects in monolayer TMDs likewise produce localized electronic perturbations and can drive charge redistribution in adjacent conducting regions~\cite{RefaelyAbramson2018_PRL,Schuler2019_PRL,Krause2021_PRL,Bobzien2025_PRL}. Taken together, these observations raise the possibility that defects located away from the edge may still influence edge-controlled injection. The central question is therefore whether a remote defect can modify the measured SBH through nonlocal electrostatic coupling to the contact edge.

In this work, we formulate the SBH of a top contact as a nonlocal edge response. Rather than starting from a spatial distribution of local barriers, we treat the measured SBH as an edge observable driven by a remote electrostatic perturbation. Using density-functional-theory-based transport calculations for Ti--MoS$_2$ and Au--MoS$_2$, we show that this mechanism generates large, metal-dependent SBH variations and can account for a substantial part of the broad experimental spread reported for top-contacted 2D semiconductors. Within this framework, we derive the defect-to-edge response, compute the resulting SBH shifts, and identify nonlocal electrostatics mediated by edge-proximal response as a key source of SBH variability.

\begin{figure}[t]
\centering
\includegraphics[width=\columnwidth]{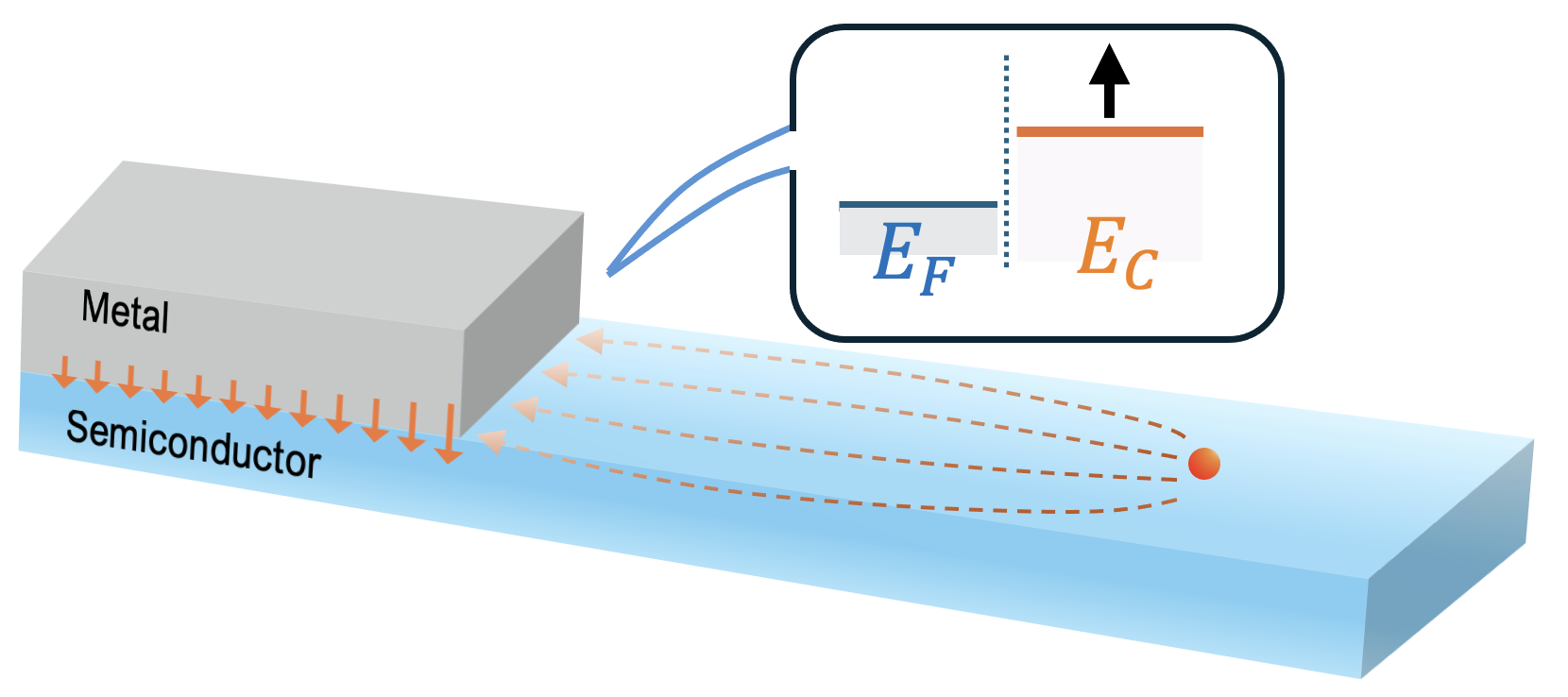}
\caption{Schematic of the nonlocal electrostatic mechanism in a top-contacted metal--semiconductor junction. A defect in the uncovered channel acts as a remote electrostatic source, and its perturbation propagates toward the contact edge. There it redistributes the edge dipole and modulates the local potential lineup, thereby changing the edge-controlled Schottky barrier height. The inset shows the resulting shift of the conduction-band edge $E_C$ relative to the metal Fermi level $E_F$.}
\label{fig:vertical_lateral_schematic}
\end{figure}

We consider a top-contact geometry with a covered region ($x<0$), an uncovered channel ($x>0$), and a contact edge at $x=0$, as sketched in Fig.~\ref{fig:vertical_lateral_schematic}. For a defect-induced electrostatic source $s_d(x)$ in the channel, the relevant quantity is the response of the edge barrier to that remote perturbation. The problem is thus to determine the defect-to-edge response kernel that maps $s_d(x)$ onto the change in the lateral Schottky barrier.

For an $n$-type lateral contact, the natural barrier observable is the conduction-band step across the contact edge, namely, the difference between the conduction-band minimum on the channel side ($E_C^{\rm ch}$) and that on the covered side ($E_C^{\rm cov}$)~\cite{Tung2001_PRB,Tung2014_APR,Wang2021_RPP},
\begin{equation}
\Phi_B^{\rm lat}=E_C^{\rm ch}(0^+)-E_C^{\rm cov}(0^-).
\label{eq:model_phi_lat}
\end{equation}
The band edge on either side may be decomposed as
\begin{equation}
E_C(x)=E_C^{(0)}-e\,u(x)+J\,\theta(x),
\label{eq:model_Ec_decomp}
\end{equation}
where $E_C^{(0)}$ is the reference conduction-band minimum, $u(x)$ is the smooth electrostatic potential, and $J$ is the contact-specific lineup offset generated by the edge dipole~\cite{Tung2000_PRL,Tung2001_PRB,Tung2014_APR}. Because $u(x)$ is continuous across the same semiconductor sheet, its direct contribution cancels in Eq.~\eqref{eq:model_phi_lat}. The barrier shift is therefore controlled by the change of the edge lineup,
\begin{equation}
\Delta\Phi_B^{\rm lat}=\Delta J.
\label{eq:model_phi_equals_J}
\end{equation}
Since the edge dipole is governed by the local semiconductor-metal potential difference, we write in linear response~\cite{Tung2001_PRB,Tung2014_APR,Gong2014_NanoLett}
\begin{equation}
\Delta J=\chi_e\bigl[u(0)-V_M\bigr],
\label{eq:model_J_linear}
\end{equation}
where $\chi_e=\partial J/\partial u(0)$ is the edge-lineup susceptibility and $V_M$ is the metal potential, taken to be constant. Equation~\eqref{eq:model_J_linear} is understood as a linearization about a reference contact configuration, with stronger short-range or nonlinear effects absorbed into effective parameters. The problem is thus reduced to the response of the edge potential $u(0)$ to a remote source.

To obtain that response, we use the standard quadratic free-energy functional of the electrostatic potential $u(x)$,
\begin{equation}
\begin{aligned}[t]
\mathcal{F}[u]
=&\frac{1}{2}\int_0^\infty dx
\left[
\kappa\left(\partial_x u\right)^2
+C_{\rm ch}u^2
\right]
+\frac{K_e}{2}u(0)^2 \\
&-\int_0^\infty dx\, s_d(x)u(x),
\end{aligned}
\label{eq:model_free_energy}
\end{equation}
where $\kappa=\varepsilon_{\parallel}t$ is the in-plane electrostatic stiffness of the channel, with $\varepsilon_{\parallel}$ the in-plane permittivity and $t$ the thickness~\cite{Berkelbach2013_PRB}, $C_{\rm ch}$ is the phenomenological restoring capacitance of the uncovered channel, analogous to the channel capacitance discussed for monolayer MoS$_2$ transistors~\cite{Bennett2023_NanoLett,Chu2014_SciRep,Velicky2016_NanoLett}, and $K_e$ is the local edge stiffness associated with the interfacial dipolar response~\cite{Tung2001_PRB,Tung2014_APR,Gong2014_NanoLett}. Minimization of Eq.~\eqref{eq:model_free_energy} yields
\begin{equation}
-\kappa u''(x)+C_{\rm ch}u(x)=s_d(x), \qquad x>0,
\label{eq:model_bulk_eq}
\end{equation}
with boundary condition $\kappa u'(0)=K_e u(0)$, expressing balance between the incoming channel field and the local edge restoring force. The solution can therefore be written as
\begin{equation}
u(0)=\int_0^\infty dx'\,W(x')\,s_d(x'),
\label{eq:model_u0_kernel}
\end{equation}
with the kernel
\begin{equation}
W(x')
=
\frac{\lambda_l}{\kappa+K_e\lambda_l}
e^{-x'/\lambda_l}.
\label{eq:model_W_kernel}
\end{equation}

Here $\lambda_l\equiv\sqrt{\kappa/C_{\rm ch}}$ is the propagation length of the defect-induced field in the uncovered channel. Substituting Eq.~\eqref{eq:model_u0_kernel} into Eq.~\eqref{eq:model_J_linear}, we obtain
\begin{equation}
\Delta\Phi_B^{\rm lat}
=
\chi_e \int_0^\infty dx'\,
\frac{\lambda_l}{\kappa+K_e\lambda_l}
e^{-x'/\lambda_l}s_d(x').
\label{eq:model_phi_general_kernel}
\end{equation}
This is the central result of the model. The lateral barrier is a nonlocal edge observable whose response to a remote source is governed by a defect-to-edge kernel. The effective range is set by $\lambda_l$, which is controlled by the competition between in-plane electrostatic spreading and restoring screening and is therefore a property of the semiconductor channel. A larger $\lambda_l$ allows defects deep inside the channel to remain coupled to the contact edge, so the measured SBH reflects a mesoscopic disorder environment rather than only the local atomic structure of the perimeter. A shorter $\lambda_l$ suppresses this long-range coupling and makes the barrier predominantly sensitive to defects located very near the edge.

The effective amplitude is set by the prefactor $\chi_e\lambda_l/(\kappa+K_e\lambda_l)$. In contrast to the range, this amplitude carries the contact-specific response through $K_e$ and $\chi_e$ and is therefore tied to the metal-contact environment. The factor $(\kappa+K_e\lambda_l)^{-1}$ determines how efficiently the incoming channel perturbation shifts the edge potential, while $\chi_e$ determines how strongly that edge-potential shift is converted into lineup modulation. Consequently, comparable defect landscapes can produce markedly different SBH shifts if the edge compliance or interfacial dipolar susceptibility differs.

\begin{figure}[!t]
\centering
\includegraphics[width=\columnwidth]{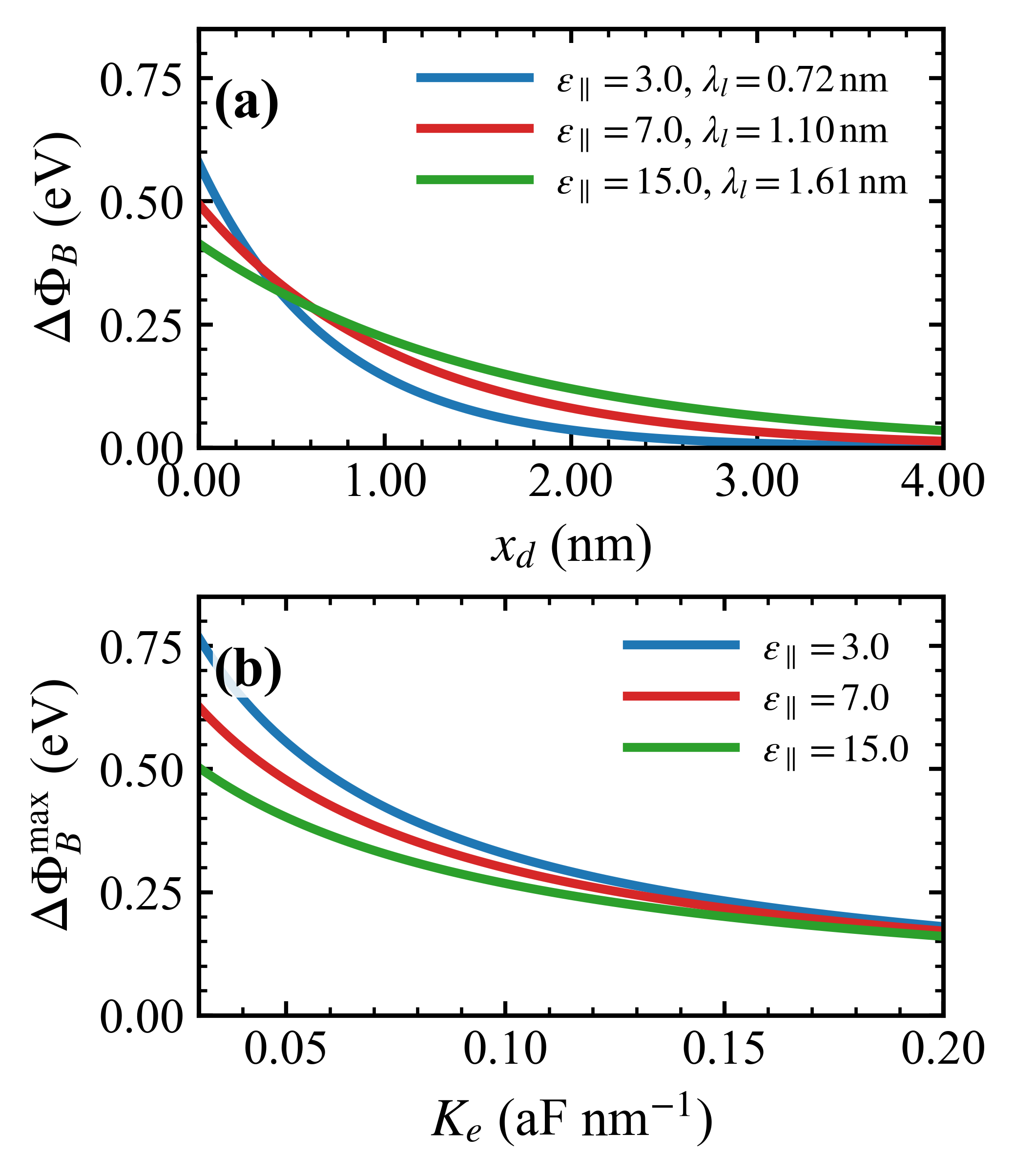}
\caption{Representative parameter dependence of the nonlocal barrier shift predicted by Eq.~\eqref{eq:model_phi_point}. Panel (a) uses $\varepsilon_{\parallel}=7$, $t=0.6\,{\rm nm}$, $\lambda_l=1.1\,{\rm nm}$, $\kappa=\epsilon_0\varepsilon_{\parallel}t=0.037\,{\rm aF}$, $C_{\rm ch}=\kappa/\lambda_l^2=3.1\,\mu{\rm F\,cm^{-2}}$, $K_e=0.047\,{\rm aF\,nm^{-1}}$, and $\chi_eS_d=4.0\times10^{-7}\,{\rm eV\,\mu F\,cm^{-1}}$. In (a), $\Delta\Phi_B^{\rm lat}(x_d)$ is shown for fixed $K_e=0.047\,{\rm aF\,nm^{-1}}$ and $\varepsilon_{\parallel}=3,\,7,\,15$. In (b), $\Delta\Phi_B^{\max}$ is shown as a function of $K_e$ for the same $\varepsilon_{\parallel}$. The adopted $\varepsilon_{\parallel}$ and $t$ are consistent with reported monolayer MoS$_2$ dielectric and structural scales~\cite{Laturia2018_npj2D,Bennett2023_NanoLett}; $C_{\rm ch}$ is obtained from $\kappa/\lambda_l^2$ and is comparable to reported monolayer MoS$_2$ transistor capacitance scales~\cite{Bennett2023_NanoLett}. The inferred $K_e$ values parameterize the partial-pinning and interfacial-dipole response known for metal--MoS$_2$ and metal--semiconductor contacts~\cite{Gong2014_NanoLett,Tung2001_PRB,Tung2014_APR}.}
\label{fig:model_parameter_panels}
\end{figure}

For a point defect, we approximate the source as $s_d(x)=S_d\delta(x-x_d)$, where $S_d$ is the effective defect charge. Equation~\eqref{eq:model_phi_general_kernel} then reduces to
\begin{equation}
\Delta\Phi_B^{\rm lat}(x_d)
=\chi_e S_d\frac{\lambda_l}{\kappa+K_e\lambda_l}e^{-x_d/\lambda_l},
\label{eq:model_phi_point}
\end{equation}

The range and amplitude of the SBH variation due to point defect are illustrated in Fig.~\ref{fig:model_parameter_panels}. Figure~\ref{fig:model_parameter_panels}(a) shows that, at fixed contact response, increasing $\varepsilon_{\parallel}$ broadens the $x_d$ dependence, so remote defects remain coupled to the edge over a longer distance; the range of the nonlocal effect is therefore set mainly by the channel electrostatics. Figure~\ref{fig:model_parameter_panels}(b) shows that increasing $K_e$ lowers $\Delta\Phi_B^{\max}$ with much weaker impact on that distance scale, indicating that the contact chiefly controls how efficiently the incoming perturbation is converted into lineup modulation. The nonlocal barrier response is thus governed by a clear division of roles: the semiconductor determines how far the defect influence propagates, whereas the contact determines how strongly that remote perturbation is expressed in the barrier.

\begin{figure*}[!t]
\centering
\includegraphics[width=\textwidth]{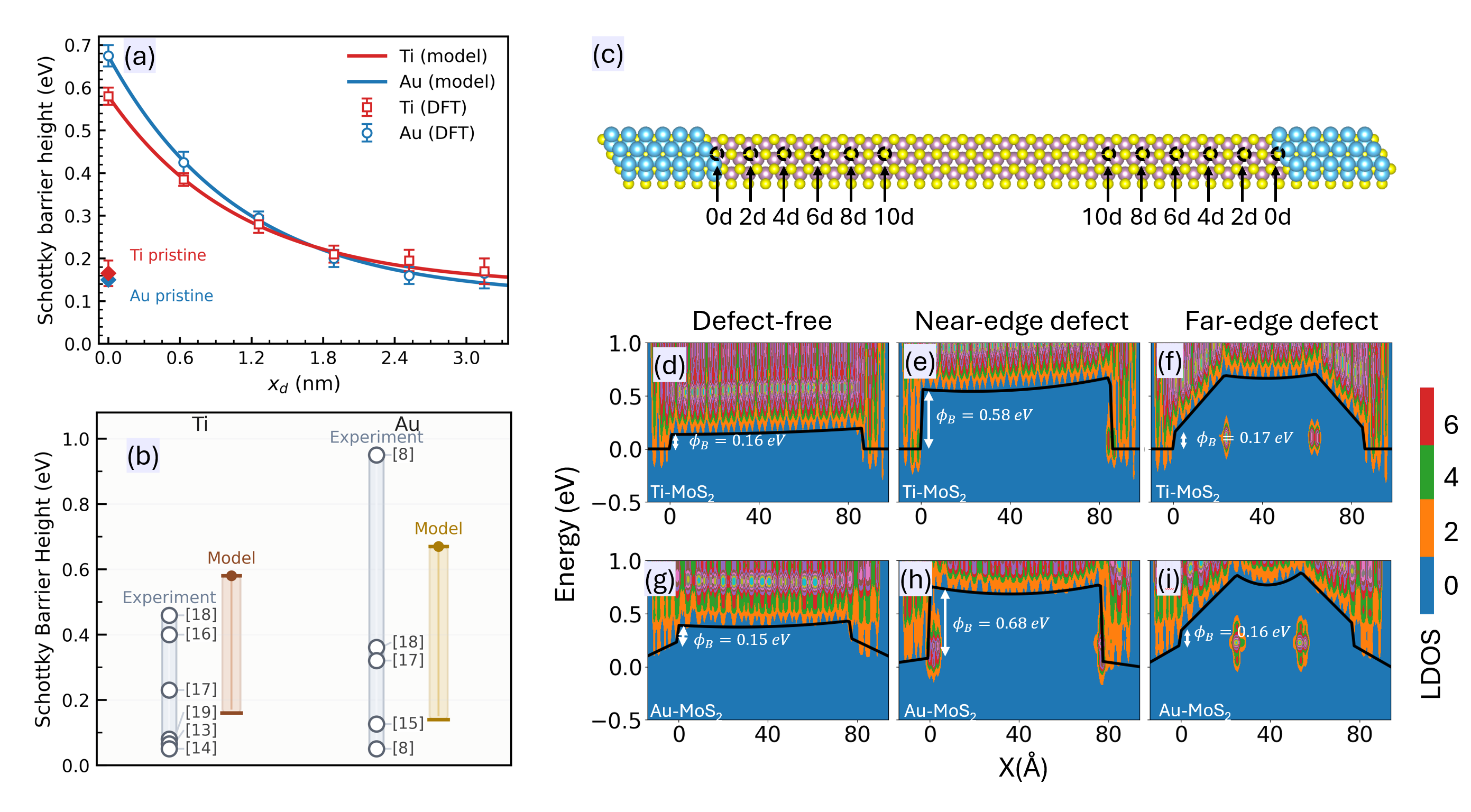}
\caption{(a) DFT-extracted lateral SBH versus defect--edge distance with electrostatic model curves. The curves use shared $\lambda_l=1.1\,{\rm nm}$, $\kappa=0.037\,{\rm aF}$, and $C_{\rm ch}=3.1\,\mu{\rm F\,cm^{-2}}$, following monolayer MoS$_2$ dielectric and thickness parameters and the overall parameter set remaining physically reasonable and consistent with MoS$_2$ capacitance scales and metal--MoS$_2$ contact-response behavior~\cite{Laturia2018_npj2D,Bennett2023_NanoLett,Chu2014_SciRep,Velicky2016_NanoLett,Gong2014_NanoLett,Tung2001_PRB,Tung2014_APR}. Error bars denote LDOS extraction uncertainty. (b) Experimental SBH benchmark for Ti--MoS$_2$ and Au--MoS$_2$, compared with the predicted SBH span. (c) Two-probe DFT--NEGF setup with sulfur vacancies placed at separations $0d$--$10d$ from the contact edge, where $d=3.15\,\text{\AA}$ and $L_{\mathrm{ch}}=11.8\,\mathrm{nm}$. (d)--(f) Atom-projected LDOS for Ti--MoS$_2$ in the pristine, near-edge-defect, and far-edge-defect cases. (g)--(i) Corresponding LDOS for Au--MoS$_2$. Black curves mark the extracted conduction-band edge and the model profile used to define the effective SBH; intermediate-distance LDOS maps are provided in the Supplemental Material.}
\label{fig:ldos_sbh_summary}
\end{figure*}

We next test the nonlocal electrostatic picture against first-principles calculations for Ti--MoS$_2$ and Au--MoS$_2$ top contacts using density-functional-theory (DFT)-based nonequilibrium Green's function (NEGF) calculations. The local density of states (LDOS) is evaluated for each defect position, from which the effective SBH is extracted from the band-edge alignment near the contact edge. To compare the model with these results, we treat $\chi_e$, $K_e$, $C_{\rm ch}$, $\varepsilon_{\parallel}$, and $s_d$ as effective phenomenological parameters constrained by established descriptions of edge-lineup response, channel electrostatics, and defect perturbations~\cite{Tung2001_PRB,Tung2014_APR,Gong2014_NanoLett,Berkelbach2013_PRB,Bennett2023_NanoLett,Chu2014_SciRep,Velicky2016_NanoLett}, and then fine-tune them to reproduce the calculated trends. This parametrization allows $\kappa$ and the resulting interaction length $\lambda_l$ to emerge from a single effective parameter set: $\lambda_l=1.1\,{\rm nm}$ and $\kappa=0.037\,{\rm aF}$ give $C_{\rm ch}=3.1\,\mu{\rm F\,cm^{-2}}$, consistent with reported MoS$_2$ capacitance scales~\cite{Bennett2023_NanoLett,Chu2014_SciRep,Velicky2016_NanoLett}, while the inferred $K_e=0.047\,{\rm aF\,nm^{-1}}$ is consistent with the strong partial pinning and spatially varying contact response reported for metal--MoS$_2$ interfaces~\cite{Kim2017_ACSNano,Bampoulis2017_ACSAMI,Gong2014_NanoLett,Zhang2022_ACSAMI}. The model--DFT comparison is shown in Fig.~3(a), the model--experimental comparison is summarized in Fig.~3(b), and the corresponding atomic structures are given in Fig.~3(c), where defects are introduced at varying distances from the contact edge.

Figure~3(a) directly validates the model against first-principles data. The DFT-extracted lateral SBHs for both metals follow the predicted defect-distance dependence over the full range using a common interaction length of $\lambda_l=1.10$ nm. As a consistency check, allowing the decay length to vary independently gives $\lambda_l^{\rm Ti}=1.12\pm0.09$ nm and $\lambda_l^{\rm Au}=1.09\pm0.07$ nm, differing from the shared value by only about 2\% and 1\%, respectively. This close agreement supports the central conclusion that $\lambda_l$ is governed primarily by channel electrostatics. Figure~3(b) places the calculated variation beside the curated experimental SBH spread. For Ti--MoS$_2$, the reported values fall within the window spanned by the pristine and defect-enhanced results, indicating that the model captures the experimentally relevant scale of variation. For Au--MoS$_2$, the experimental spread is broader, but the calculations reproduce a substantial part of that range, indicating that nonlocal defect response is a major contributor to the reported variability, with the remaining spread likely reflecting additional variations in defect density and contact formation conditions.

The DFT data in Fig.~3(a) are obtained from the relaxed structures in Fig.~3(c). First-principles calculations were performed with VASP~\cite{Kresse1996_PRB} using the projector augmented-wave method~\cite{Blochl1994_PRB} and a plane-wave cutoff of 520~eV. The Brillouin zone was sampled with a Monkhorst--Pack mesh corresponding to a reciprocal spacing of $\approx 0.04\,\text{\AA}^{-1}$, and structures were relaxed until the maximum residual force was below $20\,\text{meV}\,\text{\AA}^{-1}$. A vacuum region of at least $12\,\text{\AA}$ suppresses interactions between periodic images. Quantum transport was computed within the DFT--NEGF framework using Nanodcal~\cite{Taylor2001_PRB} with self-consistent convergence of the reduced density matrix. Exchange and correlation were treated within the Perdew--Burke--Ernzerhof generalized-gradient approximation~\cite{PBE1996_PRL} using a double-$\zeta$ plus polarization basis and a real-space mesh cutoff of 80~Hartree. The equilibrium metal--MoS$_2$ separations are $2.12\,\text{\AA}$ for Ti and $2.66\,\text{\AA}$ for Au, consistent with prior benchmarks~\cite{Pan2019_ACSANM}. To quantify the defect-position dependence, we introduce a single sulfur vacancy in the MoS$_2$ channel at separations $x_0=n d$ from the contact edge, with $n=0,2,\ldots,10$ and $d=3.15\,\text{\AA}$, while neglecting gates and substrates to isolate intrinsic contact electrostatics. The lateral SBHs in Fig.~3(a) are extracted from atom-projected LDOS maps by tracing the conduction-band edge along the transport direction and measuring the band-edge step at the contact edge.

Figure~3(d)--3(f) shows representative LDOS maps for Ti--MoS$_2$. Because the covered MoS$_2$ is metallized beneath Ti, the relevant SBH is the lateral band-edge step at the contact edge. The representative pristine and near-edge-defect cases yield $\Phi_B^{\mathrm{lat}}=0.16$ and $0.58$ eV, respectively. Figure~3(g)--3(i) shows the corresponding Au--MoS$_2$ maps. Beneath Au, a small vertical SBH remains, but the lateral barrier still dominates; the representative maps give $\Phi_B^{\mathrm{lat}}=0.68$ eV for a near-edge defect and $0.16$ eV at large separation. Together with the intermediate-distance calculations provided in the Supplemental Material, these LDOS extractions generate the Ti and Au curves in Fig.~3(a) and yield a nonlocal defect--edge interaction length of $\lambda_l=1.1$ nm.

The finite interaction length extracted here establishes nonlocal electrostatics as a central factor governing Schottky barriers in top-contacted 2D semiconductors, beyond the conventional local interface picture. By linking barrier formation to edge-mediated coupling with nearby defects over nanometer length scales, this framework helps explain longstanding inconsistencies in reported SBHs and provides a basis for predictive, spatially engineered contact design in atomically thin devices.
We have shown that, in top-contacted 2D semiconductors, the measured Schottky barrier is a nonlocal, edge-controlled quantity rather than a purely local interface property. Because carrier injection is governed by the lateral barrier at the contact edge, defects within a finite defect--edge interaction length can strongly modulate the effective SBH even when they are not located directly at the edge. This framework resolves a long-standing source of variability in the 2D-contact literature by showing how widely different SBHs can arise naturally from the combined effects of edge electrostatics and defect position in nominally similar metal--semiconductor contacts. More broadly, it identifies the edge environment and the spatial distribution of defects near the contact as key design variables for engineering carrier injection in 2D devices. The present mechanism therefore establishes nonlocal defect-induced modulation of the edge lineup as a distinct source of barrier variability in 2D top contacts, beyond classical inhomogeneous-Schottky and pinch-off pictures.

\begin{suppinfo}
Additional computational details, LDOS maps, and supporting analysis (DOCX).
\end{suppinfo}

\begin{acknowledgement}
We gratefully acknowledge support from Singapore NRF T-CRP (Grant No. T-CRP-2025-0030).
\end{acknowledgement}

\end{document}